\documentclass[showpacs,prd,10pt,twocolumn,superscriptaddress,nofootinbib,aps]{revtex4-1}

\usepackage{ae}
\usepackage{bm} 
\usepackage{xcolor}
\usepackage{amsmath}
\usepackage{amssymb}
\usepackage{amsfonts}
\usepackage{graphicx}
\usepackage{slashed}
\usepackage{wasysym}
\usepackage{units}
\usepackage{hyperref}

\newcommand{\Eqref}[1]{Eq.~\eqref{#1}}

\newcommand{\bk}{\bar{\kappa}}
\newcommand{\bms}{\bar{m}}
\newcommand{\be}{\bar{e}}
\newcommand{\Gprop}{G}
\newcommand{\Nf}{N_{\text{f}}}

\usepackage[utf8]{inputenc}

\makeatletter
\def\fps@figure{ht}
\def\fps@table{ht}
\makeatother

\allowdisplaybreaks

\begin{document}

\setlength{\unitlength}{1mm}

\title{Asymptotically Safe QED}

\author{Holger Gies}
\affiliation{Theoretisch-Physikalisches Institut, Abbe Center of Photonics, Friedrich-Schiller-Universit\"at Jena, Max-Wien-Platz 1, 07743 Jena, Germany}
\affiliation{Helmholtz-Institut Jena, Fr\"obelstieg 3, 07743 Jena, Germany}
\author{Jobst Ziebell}
\affiliation{Theoretisch-Physikalisches Institut, Abbe Center of Photonics, Friedrich-Schiller-Universit\"at Jena, Max-Wien-Platz 1, 07743 Jena, Germany}

\date{\today}

\begin{abstract}
  High-energy completeness of quantum electrodynamics (QED) can be
  induced by an interacting ultraviolet fixed point of the
  renormalization flow. We provide evidence for the existence of two
  of such fixed points in the subspace spanned by the gauge coupling,
  the electron mass and the Pauli spin-field coupling. Renormalization
  group trajectories emanating from these fixed points correspond to
  asymptotically safe theories that are free from the Landau pole
  problem. We analyze the resulting universality classes defined by
  the fixed points, determine the corresponding critical exponents,
  study the resulting phase diagram, and quantify the stability of our
  results with respect to a systematic expansion scheme. We also
  compute high-energy complete flows towards the long-range
  physics. We observe the existence of a renormalization group
  trajectory that interconnects one of the interacting fixed points
  with the physical low-energy behavior of QED as measured in
  experiment.  Within pure QED, we estimate the crossover from
  perturbative QED to the asymptotically safe fixed point regime to
  occur somewhat below the electroweak scale.
\end{abstract}

\maketitle

\section{Introduction}
\label{sec:intro}

While quantum electrodynamics (QED) is the most precisely tested part
of the (quantum field) theory of the standard model of particle
physics (see, e.g., \cite{Hanneke:2008tm}), it has been known early on
that its perturbative structure is plagued by a singularity in the
running coupling at finite scales, the so-called Landau pole
\cite{Landau:1954,Landau:1955}. Attempts to search for a cure for this
consistency problem in the nonperturbative strong-coupling domain also
date back to the early days of quantum field theory
\cite{GellMann:1954fq,Baker:1969an,Johnson:1973pd}. In absence of a
convincing solution, QED is considered to be a ``trivial'' theory, in
the sense that the theory is assumed to be a consistent quantum field
theory only at the prize of having no interactions.

In fact, evidence for triviality has been provided by lattice simulations
\cite{Gockeler:1997dn,Kim:2000rr,Kim:2001am} as well as nonperturbative functional methods
\cite{Gies:2004hy}, though the resulting picture is more involved (and
different from the triviality arising, e.g., in $\phi^4$ theory
\cite{Luscher:1987ek,Luscher:1988uq,Hasenfratz:1987eh}): if QED was in a strong-coupling regime at a high-energy
scale $\Lambda$, interactions would trigger chiral symmetry breaking
\cite{Miransky:1984ef,Aoki:1996fh} much in the same way as in QCD. As a consequence, such a
strong-coupling realization of QED would go along with electron masses of the
order of the high scale $m\sim \Lambda$ in contradistinction to the observed
small mass of the electron and its approximate chiral symmetry in comparison
with generic standard-model scales. Therefore, the Landau pole representing a
strong-coupling regime is not connected by a line of constant physics with QED
as observed in Nature \cite{Gockeler:1997dn}; nevertheless, the existence of a
chiral-symmetry breaking phase imposes a scale $\Lambda_{\text{max}}$ up to
which QED as an effective field theory can be maximally extended \cite{Gies:2004hy}. In pure QED,
this scale has been estimated as $\Lambda_{\text{max,QED}}\simeq
10^{278}$GeV. As far as an ultraviolet (UV) completion of QED is concerned,
the conclusion is similar to that of naive perturbation theory: a simple
high-energy completion of QED does not seem to exist.

From the modern perspective of the standard model, QED is merely the
low-energy remnant of the electroweak sector of the standard model as a
consequence of the Brout-Englert-Higgs mechanism. However, the
hypercharge U(1) factor of the gauge group of the SM exhibits a high energy
renormalization group (RG) behavior qualitatively similar to QED; the
high-energy location of the corresponding Landau pole of perturbation theory
suggests the existence of a scale of maximum UV extent of the standard model
of $\Lambda_{\text{max,SM}}\simeq 10^{40}$GeV. It is fair to
say that the physical relevance of such a scale remains unclear, since it is
much larger than the Planck scale where the renormalization behavior of the
particle physics sector is expected to be modified by quantum gravitational
effects. Still, this problem appears to be generic for models with U(1)
factors; in fact the Landau pole typically moves to smaller scales for new
physics models with a larger sector of U(1)-charged scalar or fermionic
particles and can thus easily drop below the Planck scale.

Within QED-like (asymptotically non-free) theories, analytic
properties of the 't Hooft expansion at large $\Nf$
\cite{PalanquesMestre:1983zy,Gracey:1996he} have been used in combination with
high-order perturbation theory to actively search for UV fixed points
\cite{Shrock:2013cca}, and are currently studied with
renewed interest using novel resummation techniques
\cite{Antipin:2017ebo, Antipin:2018zdg,Dondi:2020qfj, Dondi:2019ivp},
aiming at addressing the fate of these theories in the deep 
UV. Proposed solutions of the problem of high-energy incompleteness
caused by a U(1) factor typically go much beyond the particle content
of pure QED-like systems. One example is given by asymptotically safe
particle physics models \cite{Litim:2014uca} which require a large
number of additional vector-like fermions \cite{Mann:2017wzh} but go
along with a nonperturbative scalar sector \cite{Pelaggi:2017abg}; the
mechanisms that help controlling UV fixed points in non-abelian gauge-Yukawa
models have recently been shown to be, in principle, also available in
corresponding abelian systems but definite answers require a non-perturbative
analysis \cite{Held:2020kze}. A
natural solution might be given by an embedding of the U(1) factor
into a unified non-abelian group, provided that a suitable physical
spectrum arises \cite{Maas:2017xzh,Sondenheimer:2019idq}. A rather
interesting possibility has been discussed within the combined system of QED
and gravitational fluctuations based on the asymptotic safety scenario
of quantum gravity
\cite{Harst:2011zx,Christiansen:2017gtg,Eichhorn:2017lry,Eichhorn:2017muy,Eichhorn:2019dhg},
since the combined system can develop a UV fixed point, for which the
low-energy QED coupling becomes a predictable quantity.

Returning to a pure QED perspective, it has recently been observed within an
effective-field theory analysis that a finite Pauli term (the spin-field
coupling) can be sufficient to screen the perturbative Landau pole
\cite{Djukanovic:2017thn} and render the minimal gauge coupling finite. Within
the effective field theory paradigm, this suggests that QED triviality could
be an artifact of truncating the effective field theory at leading order. If
so, high-energy completion would still require an embedding into a
``new-physics'' framework which remains unknown at this point.

In the present paper, we explore the possibility whether QED could be
asymptotically safe in a theory space larger than what has so far been
considered in lattice simulations or functional methods. Inspired by
\cite{Djukanovic:2017thn}, we include the Pauli coupling $\kappa$
parametrizing the unique dimension-5 operator to lowest-derivative order and thus a next-to-leading order
term in an operator expansion of the effective action. A reason to disregard
this term in earlier studies might have been given by the fact that the Pauli
term breaks chiral symmetry explicitly (apart from perturbative
non-renormalizability). By contrast, high-energy studies typically assume
asymptotic symmetry \cite{Lee:1974gua}, as the electron mass being the source
of chiral symmetry breaking (in pure QED) is implicitly assumed to be
irrelevant in comparison to all other momentum scales at
high energies. Counterexamples to this scenario have been constructed only
recently in the context of non-abelian Higgs-(Yukawa) models 
\cite{Gies:2013pma,Gies:2015lia,Gies:2016kkk,Gies:2018vwk,Gies:2019nij},
exhibiting mass scales that grow proportionally to an (RG) scale; see
\cite{Gies:2009hq,Gies:2009sv} for earlier toy-model examples.

In fact, using modern functional renormalization group techniques, we find
evidence for the existence of interacting RG fixed points in the theory space
spanned also by the Pauli coupling. RG trajectories that emanate from such
fixed points correspond to high-energy complete realizations of QED with a
fixed set of physical parameters and a full predictive power for the
long-range behavior of the theory. The dynamics induced by the Pauli coupling
exhibits several interesting features: For increasing Pauli coupling, its RG
flow turns from irrelevant to relevant, i.e., the power-counting scaling is
compensated by quantum fluctuations. Also the running of the gauge
coupling $e$ is driven towards asymptotic freedom (whereas $\kappa$ is
asymptotically safe). We observe several fixed points that qualitatively
differ by the presence or absence of a finite value for the electron mass
(measured in units of the RG scale), by the number of relevant directions
corresponding to the number of physical parameters, and by the properties of
the long-range physics. We identify RG trajectories that interconnect the
physical values for the low-energy parameters of real QED with one of the UV
fixed points, thereby constructing a high-energy complete version of QED with
only photon and electron degrees of freedom.

Our paper is structured as follows: In Sect.~\ref{sec:model}, we introduce the
subspace of the QED theory space to be screened for the existence of fixed
points. Sect.~\ref{sec:floweq} presents our results for the RG flow equation
in that subspace. In Sect.~\ref{sec:FP}, we present the results of our RG
fixed points search and classify the resulting universality
classes. Sect.~\ref{sec:flowIR} is devoted to a construction of UV complete
trajectories and an analysis of the resulting long-range properties. In
Sect.~\ref{sec:conc}, we conclude and discuss possible implications of our
results for pure QED in the context of an embedding into a standard-model like
theory. Further technical details are summarized in the appendix.

\section{QED with a Pauli term}
\label{sec:model}

Let us consider pure QED, consisting of an electromagnetic U(1) gauge field
$A_\mu$ interacting with a massive electron that is described by a Dirac
spinor $\psi$. In addition to the standard kinetic terms, the mass term, and
the minimal coupling, we also consider a Pauli term, parametrizing the
(anomalous) coupling of the electron to the electromagnetic field. Using
Euclidean spacetime and Dirac-space conventions, the bare action reads
\begin{equation}
  S=\int_x \bar{\psi} i\slashed{D}[A] \psi +\frac{1}{4} F_{\mu\nu} F^{\mu\nu} - i\bms \bar{\psi} \psi + i \bk
  \bar{\psi} \sigma_{\mu\nu} F^{\mu\nu} \psi \label{eq:bareS},
\end{equation}
with $D_\mu[A]=\partial_\mu - i\be A_\mu$ denoting the covariant derivative,
and $\bms$, $\be$, $\bk$ representing the bare mass and couplings.
Note that the factors of $i$ in front of the mass term and the Pauli spin
coupling arise from the Euclidean description; the action satisfies
Osterwalder-Schrader reflection positivity. The action features a local U(1)
(vector) gauge invariance. In addition to the electron mass term $\sim \bms$, also the Pauli
coupling $\sim \bk$ breaks the global chiral symmetry explicitly.

Let us briefly sketch the reasoning behind a conventional perturbative RG
analysis: Based on the assumption that the theory is close to the Gaußian
fixed point at an initial high-energy scale $\Lambda$ with all couplings
$\be,\bk,\dots \lesssim\mathcal{O}(1)$ (suitably measured in units of
$\Lambda$), the Pauli term as a dimension-5 operator as well as all
possible higher order couplings $\bk, \dots$ are expected to be governed by
their power-counting dimension (possibly amended by logarithmic
corrections). As a result, the Pauli coupling is expected to scale as $\bk
\sim k/\Lambda$ towards lower scales $k\ll \Lambda$, exhibiting RG irrelevance
for the long-range physics. Higher-order operators are expected to be correspondingly
power-suppressed. By contrast, the dimensionless RG-marginal gauge coupling
runs logarithmically, as is captured by the $\beta$ function for the suitably
renormalized coupling $e$ (see below),
\begin{equation}
  \beta_e= k \frac{de}{dk}= \frac{e^3}{12\pi^2} + \mathcal{O}(e^5). \label{eq:pertbetae}
\end{equation}
The running of the coupling obtained from the integrated $\beta_e$ function
exhibits a logarithmic decrease of the coupling towards lower scales $k$ and a
Landau-pole singularity towards the UV; the latter signals the break-down of
the perturbative reasoning towards higher energies.

In this discussion, we have already implicitly assumed the mass to be smaller
than any of the scales $k,\Lambda$ (or loop momenta). This assumption
characterizes the \textit{deep Euclidean region} where a possible finite mass
can be ignored. The finiteness of the renormalized mass $m$ then only becomes relevant at low
scales $k\sim m$, where threshold effects lead to a decoupling of
massive particles from the
flow.

Upon embedding the pure QED sector into the standard model, the corresponding mass term
arises from the Higgs mechanism and is seeded by the Yukawa coupling to the
Higgs field. The latter is an RG-marginal coupling as well and preserves
chiral symmetry. This, together with the assumption of asymptotic symmetry
\cite{Lee:1974gua} justifies the procedure to ignore particle masses in the high-energy
analysis of standard-model like theories.

This work is devoted to an analysis of the nonperturbative RG flow in pure QED
theory space including the Pauli term. The anticipated existence of an
interacting RG fixed point can invalidate simple power-counting arguments for
the Pauli coupling. If so, high-energy quantum fluctuations could render the
Pauli term RG relevant and exert a strong influence on the high-energy
behavior of the gauge coupling. In addition to strong-coupling effects, explicit
chiral symmetry breaking triggered by the Pauli term makes it necessary to
consider the flow of the mass term on the same footing as the couplings. Since
a finite mass term can generically induce decoupling, it remains a nontrivial
question as to whether RG trajectories exist along which the high-energy
behavior can be separated from the physical low-energy electron mass scale.

\section{QED flow equations}
\label{sec:floweq}

Whereas the action in \Eqref{eq:bareS} could be straightforwardly treated with
standard effective field theory methods in the deep Euclidean region, the fact
that the Pauli term breaks chiral symmetry suggests to use a formalism where
all sources of symmetry breaking including the mass term are treated on the
same footing. In order to study the RG flow beyond the bias of the deep
Euclidean region, we use the functional RG formulated in terms of the Wetterich equation \cite{Wetterich:1992yh}
\begin{equation}
\partial_t \Gamma_k = \frac{1}{2} \mathrm{STr} \left[ \left( \partial_t R_k \right) \left( \Gamma^{(2)}_k + R_k \right)^{-1} \right],
\label{eq:Wetterich}
\end{equation}
where $t = \ln \left( k / \Lambda \right)$ denotes ``RG time'' defined in
terms of a scale $k$, separating the modes with momenta $\lesssim k$ to be
integrated out from those with momenta $\gtrsim k$ already integrated
out. This mode separation is technically implemented by a momentum-dependent
regulator $R_k(p^2)$. The resulting effective average action $\Gamma_k$
governs the dynamics of the low-momentum modes, is initiated at the high scale
by the bare action, $\Gamma_{k\to\Lambda}\to S$, and approaches the full
quantum effective action for $k\to0$, for reviews see \cite{Berges:2000ew,Pawlowski:2005xe,Gies:2006wv,Delamotte:2007pf,Braun:2011pp,Nagy:2012ef}.
In the present work, we study the flow of the system in a truncated theory
space spanned by the action 
\begin{equation}
\begin{aligned}
\Gamma_k &=\int_x\Big[ \bar{\psi} \left( i Z_\psi \slashed{\partial} + \be \slashed{A} - i\bms + i \bk
\sigma_{\mu\nu} F^{\mu\nu} \right) \psi\\
&\quad+ \frac{1}{4} Z_A F_{\mu\nu} F^{\mu\nu} + \frac{Z_A}{2 \xi} \left(
  \partial_\mu A^\mu \right)^2\Big].
\label{eq:Gammak}
\end{aligned}
\end{equation}
Here, the wave function normalizations $Z_\psi$ and $Z_A$ as well as all
couplings and mass parameters are considered as $k$ dependent. While the gauge
parameter $\xi$ could also be studied as a $k$-dependent parameter, we choose
the Landau gauge $\xi=0$ in practice, as it is a fixed point of the RG flow
\cite{Ellwanger:1995qf,Litim:1998qi}. This truncation is complete to lowest order in a derivative expansion (1st order for fermions, 2nd order for photons) and to dimension-5 in a power-counting operator expansion. A next order in derivatives would include the operators $\bar{\psi} \slashed{D} \slashed{D} \psi$ (dimension-5), $F_{\mu\nu}\square F^{\mu\nu}$ (dimension-6); a next order in the operator expansion includes four-fermion terms as studied in \cite{Aoki:1996fh,Gies:2003dp,Gies:2004hy}. 

It is convenient to express the RG flow in terms of dimensionless and
renormalized parameters: 
\begin{equation}
e = \frac{k^{\frac{d}{2}-2}\be}{Z_\psi\sqrt{Z_A} }, \quad
\kappa =  \frac{k^{\frac{d}{2}-1}\bk}{Z_\psi \sqrt{Z_A}}, \quad
m = \frac{\bms}{Z_\psi k}.
\end{equation}
and calculate their $\beta$ functions in the Landau gauge $\xi \to 0$ using equation \eqref{eq:Wetterich}.
For this, we use standard methods for the operator expansion of the Wetterich
equation \cite{Berges:2000ew,Gies:2001nw,Gies:2006wv,Braun:2011pp} in order to
project onto the operators of \Eqref{eq:Gammak}, and employ \textsc{FeynCalc}
\cite{Mertig:1990an,Shtabovenko:2016sxi} for some of the tensor
manipulations.
The results for these $\beta$ functions are rather involved as a result of the
absence of chiral symmetry and the possible finiteness of the mass term. For
generality, we list the results for a generic spacetime dimension $d$:
\begin{widetext}
\begin{eqnarray}
  \partial_t e &=&
                   e \left( \frac{d}{2}-2+ \eta_\psi + \frac{\eta_A}{2}  \right) 
 - 4 v_d \frac{\left(d-4\right)\left(d-1\right)}{d} e^3 \; l_d^{(1,\mathrm{B},\tilde{\mathrm{F}}^2)}(0,m^2) 
 - 16 v_d \frac{\left(d-2\right)\left(d-1\right)}{d} e \kappa^2 \; l_d^{(2,\mathrm{B},\tilde{\mathrm{F}}^2)}(0,m^2) \nonumber\\
&& - 32 v_d \frac{d-1}{d} e^2 \kappa m \; l_d^{(1,\mathrm{B},\mathrm{F},\tilde{\mathrm{F}})}(0,m^2,m^2) 
 - 4 v_d \frac{\left(d-2\right)\left(d-1\right)}{d} e^3 m^2 \; l_d^{(\mathrm{B},\mathrm{F}^2)}(0,m^2)\nonumber\\
&&
   - 16 v_d \frac{\left(d-4\right)\left(d-1\right)}{d} e \kappa^2 m^2 \; l_d^{(2,\mathrm{B},\mathrm{F}^2)}(0,m^2),\label{eq:dte}
\end{eqnarray}
\begin{eqnarray}
  \partial_t \kappa
  &=&
\kappa \left(\frac{d}{2}-1 + \eta_\psi + \frac{\eta_A}{2}\!   \right)
      + 16 v_d \frac{\left(d-4\right)\left(d-1\right)}{d} \kappa^3 \; l_d^{(2,\mathrm{B},\tilde{\mathrm{F}}^2)}(0,m^2)
      - 4 v_d \left( 3 \frac{\left(d-6\right)\left(d-2\right)}{d} + 1 \right) e^2 \kappa \; l_d^{(1,\mathrm{B},\tilde{\mathrm{F}}^2)}(0,m^2)
\nonumber\\
  && + 4 v_d\,  e^3 m \;\left[\frac{d-3}{d} [ l_d^{(1,\mathrm{B},\tilde{\mathrm{F}}_1,\mathrm{F})}(0,m^2,m^2)
-l_d^{(1,\mathrm{B},\mathrm{F}_1,\tilde{\mathrm{F}})}(0,m^2,m^2) ]
     -  \frac{\left(d-4\right)\left(d-1\right)}{2 d}
     l_d^{(\mathrm{B},\mathrm{F},\tilde{\mathrm{F}})}(0,m^2,m^2) 
     \right]   \nonumber\\
&& + 16 v_d\; e \kappa^2 m\; \Big[\frac{5\left(d-4\right)\left(d-3\right)}{2d}  \;l_d^{(1,\mathrm{B},\mathrm{F},\tilde{\mathrm{F}})}(0,m^2,m^2) 
 + \frac{d-3}{d} \;  l_d^{(2,\mathrm{B},\mathrm{F},\tilde{\mathrm{F}}_1)}(0,m^2,m^2)
 - \frac{d-3}{d} \;  l_d^{(2,\mathrm{B},\mathrm{F}_1,\tilde{\mathrm{F}})}(0,m^2,m^2) \nonumber\\
&&\phantom{+ 16 v_d\; e \kappa^2 m\; \Big[}-  \frac{d+2}{d} \; l_d^{(1,\mathrm{B},\mathrm{F},\tilde{\mathrm{F}})}(0,m^2,m^2)
   \Big]
   \nonumber\\
&& + 16 v_d \left( 1 - \frac{\left(d-4\right)^2}{d} \right) \kappa^3 m^2 \;   l_d^{(1,\mathrm{B},\mathrm{F}^2)}(0,m^2)
   + 4 v_d \frac{\left(d-4\right)\left(d-1\right)}{d} e^2 \kappa m^2 \; l_d^{(\mathrm{B},\mathrm{F}^2)}(0,m^2)
 , \label{eq:dtkap}
\end{eqnarray}

\begin{equation}
  \partial_t m
  =
      -m \left( 1-\eta_\psi\right)
      - 16 v_d \left( d - 1 \right) e \kappa \; l_d^{(1,\mathrm{B},\tilde{\mathrm{F}})}(0,m^2)
      + 16 v_d \left( d - 1 \right) m \kappa^2 \; l_d^{(1,\mathrm{B},\mathrm{F})}(0,m^2)
      - 4 v_d \left( d - 1 \right) e^2 m \; l_d^{(\mathrm{B},\mathrm{F})}(0,m^2).
      \label{eq:dtm}
\end{equation}
\end{widetext}
Here, we have defined $v_d=[2^{d+1}\pi^{d/2} \Gamma(d/2)]^{-1}$, and the
functions $l_d^{\cdots}(\cdots)$ parametrize threshold effects arising from
the massive fermion propagator; in addition to the explicitly highlighted mass dependence, they can also depend on the anomalous dimensions  $\eta_\psi, \eta_A$ as a consequence of ``RG improvement'' implementing a resummation of large classes of diagrams,
\begin{equation}
  \eta_A=- \partial_t \ln Z_A, \quad   \eta_\psi=- \partial_t \ln
  Z_\psi. \label{eq:eta}
\end{equation}
The threshold functions approach a finite non-negative constant for
$m, \eta_\psi, \eta_A\to0$, and vanish for $m\to\infty$ manifesting the decoupling of massive
fermion modes. As we encounter threshold functions that go beyond those
tabulated in the literature as a consequence of the absence of chiral symmetry
as well as the presence of a momentum dependent vertex, we have introduced a
new systematic notation here, which we explain in detail in the Appendix.

These flow equations are autonomous coupled ordinary differential equations
which depend on the anomalous dimensions of the fields. The latter are determined by the flow of the kinetic terms, yielding algebraic equations of the form
\begin{widetext}
\begin{eqnarray}
\eta_\psi &=& 
4 v_d \frac{ \left( d - 2 \right) \left( d - 1 \right)}{d}  e^2 \; l_d^{(\mathrm{B},\tilde{\mathrm{F}})}(0,m^2) 
 - 8 v_d \frac{d - 1}{d} e^2 \; l_d^{(1,\mathrm{B},\tilde{\mathrm{F}}_1)}(0,m^2) 
 + 16 v_d \frac{\left( d-4 \right) \left( d - 1 \right)}{d} \kappa^2 \; l_d^{(1,\mathrm{B},\tilde{\mathrm{F}})}(0,m^2) \nonumber\\
&& - 32 v_d \frac{d - 1}{d} \kappa^2 \; l_d^{(2,\mathrm{B},\tilde{\mathrm{F}}_1)}(0,m^2) 
   + 32 v_d \frac{d-1}{d} e \kappa m \; l_d^{(1,\mathrm{B},\mathrm{F}_1)}(0,m^2),
   \label{eq:etapsi}\\
%
  \eta_A &=&
             8 v_d \frac{d_\gamma \Nf}{d + 2} e^2 \; l_d^{(2,\tilde{\mathrm{F}}_1^2)}(m^2)
             + 16 v_d d_\gamma \Nf \kappa^2 m^2 l_d^{(\mathrm{F}^2)}(m^2) 
             - 16 v_d \frac{d-4}{d} d_\gamma \Nf \kappa^2 \;l_d^{(1,\tilde{\mathrm{F}}^2)}(m^2) \nonumber\\
&& - 64 v_d \frac{d_\gamma \Nf}{d} e \kappa m \; l_d^{(1,\tilde{\mathrm{F}}, \mathrm{F}_1)}(m^2,m^2) 
   + 8 v_d \frac{d_\gamma \Nf}{d} e^2 m^2 \; l_d^{(1,\mathrm{F}_1^2)}(m^2),
   \label{eq:etaA}
\end{eqnarray}
\end{widetext}
where $d_\gamma$ denotes the dimensionality of the representation of the Dirac algebra ($d_\gamma=4$ in physical QED), and $\Nf$ is the number of Dirac fermion flavors. For an understanding of the coupling dependence of the flows, the following two discrete $\mathbb{Z}_2$ symmetries are relevant: We observe that the action \eqref{eq:Gammak} is invariant under a simultaneous discrete axial transformation $\psi \to e^{i\frac{\pi}{2} \gamma_5} \psi$, $\bar\psi \to \bar\psi e^{i\frac{\pi}{2} \gamma_5}$ and a sign flip of $\bk \to -\bk$ and $\bms\to -\bms$. This $\mathbb{Z}_2$ symmetry is also visible in all flow equations and anomalous dimensions, as they remain invariant under a simultaneous sign flip of $\kappa$ and $m$.
Furthermore, charge conjugation on the level of couplings is represented by simultaneous sign flip of $e$ and $\kappa$ which is also an invariance of all $\beta$ functions and anomalous dimensions. 

While each term in the flow equations reflects the one-loop structure of the
Wetterich equation -- visible in terms of the explicitly highlighted
polynomial coupling dependence -- the flows still exhibits various
nonperturbative features: the flow of $\kappa$ coupling to the $e$ and
$m$ equations effectively corresponds to feeding back higher-order diagrams,
the anomalous dimensions in the threshold functions also yield higher-order
resummations, and the dependence of the threshold functions on the running
mass is also a nonperturbative effect. As a simple check, it is straightforward
to rediscover the perturbative limit. For this, we drop all $\kappa$ terms and
take the deep Euclidean limit $m\to 0$. In the flow equation \eqref{eq:dte}
for $e$ only the anomalous dimensions in the first scaling term remain in this
limit. Further, we observe that $\eta_\psi\to 0$ in this limit as the
seemingly remaining terms $\sim e^2$ cancel by virtue of properties of the
threshold functions. This is in agreement with the standard perturbative
result in the Landau gauge. The only non-trivial term in the flow is carried
by the anomalous dimension $\eta_A$ of the photon, finally leading to
\Eqref{eq:pertbetae} to lowest order in the coupling. 

\section{QED fixed points and universality classes}
\label{sec:FP}

The scenario of asymptotic safety relies on the existence of an interacting
non-Gaußian fixed point of the renormalization group
\cite{Weinberg:1976xy,Weinberg:1980gg}. Summarizing all dimensionless
couplings including the mass parameter into a vector $\mathbf{g}$, with
$\mathbf{g}=(e,\kappa,m)$ in the present case, a fixed point satisfies
$\partial_t\mathbf{g}|_{\mathbf{g}=\mathbf{g}^\ast}=0$, realizing the concept
of (quantum) scale invariance. In the vicinity of a
fixed point, the RG flow to linear order is governed by the properties of the 
stability matrix $\mathbf{B}_{ij}$, the eigenvalues of which are
related to the RG critical exponents $\theta_I$,
\begin{equation}
  \mathbf{B}_{ij}= \frac{\partial}{\partial g_j} \partial_t
  g_i\Bigg|_{\mathbf{g}=\mathbf{g}^\ast}, \quad \theta_I = - \text{eig}\,
  \mathbf{B}.
\end{equation}
The number of positive critical exponents corresponds to the number of RG
relevant directions. (A zero eigenvalue $\theta_I=0$ corresponds to an RG
marginal direction with higher-orders beyond the linearized regime deciding
about marginal relevance or irrelevance.) The number of relevant and
marginally relevant directions counts the number of physical parameters that
need to be fixed in order to predict the long-range behavior of the theory. In
the presence of several fixed points, each fixed point defines a universality
class: for all RG flow trajectories passing through the vicinity of the fixed
point, the long-range behavior is universally governed by these
(marginally) relevant directions. Those RG trajectories that emanate from a
fixed point are UV complete: a theory can be extended to arbitrarily high
scales with the long-range physics remaining fixed; such a trajectory defines
a ``line of constant physics''. For further details or reviews of asymptotic
safety, see \cite{Weinberg:1980gg,Percacci:2017fkn,Reuter:2019byg,Braun:2010tt,Braun:2011pp,Nagy:2012ef}.

The previously mentioned $\mathbb{Z}_2$ symmetries of the flows translate to
relations among possible fixed points of the RG flow: given any fixed point
$(e^*, \kappa^*, m^*)$, we can construct the following set of points which are
also fixed points of the RG, describing one and the same universality class
   %
%
\begin{equation}
 (-e^*, -\kappa^*, m^*),
  (e^*, -\kappa^*, -m^*),(-e^*, \kappa^*, -m^*). \label{eq:Z2}
\end{equation}

For the concrete evaluation of the flows and the search for fixed points, we concentrate on the relevant case of four spacetime dimensions $d=4$, the irreducible representation of the Dirac algebra $d_\gamma=4$ and a single fermion flavor $\Nf=1$. For simplicity, we use the the linear regulator \cite{Litim:2000ci,Litim:2001up} for which all threshold functions can be evaluated analytically, yielding rational functions of the mass arguments, see the Appendix.

As an internal consistency check, we define the leading-order (LO) evaluation
of our flows in terms of ignoring the dependence of the threshold functions on
the anomalous dimensions; i.e., we drop the higher-loop RG improvement
provided by these resummations, but keep the anomalous dimensions in the
scaling terms as they contribute to leading-loop level. If our truncation is
reliable, we expect the LO results to agree qualitatively and
semi-quantitatively with those of the full truncation for the following
reasons: First, the size of the anomalous dimensions can be viewed as a
measure for the validity of the derivative expansion as the anomalous
dimensions quantify the running of the kinetic (derivative) terms. Second,
anomalous dimensions also quantify the deviations from canonical scaling;
hence, if the anomalous dimensions are sufficiently small, higher-order
operators can be expected to remain RG irrelevant. As a self-consistency
criterion, we thus require the anomalous dimensions to be sufficiently small,
$|\eta_{A,\psi}|\lesssim \mathcal{O}(1)$. In this LO approximation, we find
the fixed points displayed in Table \ref{tbl:LOFixedPoints}. 

\begin{table}
	\centering
	\begin{tabular}{ccccccccc}
		 & $e^\ast$ & $\kappa^\ast$ & $m^\ast$ & multiplicity & $n_{\mathrm{phys}}$ & $\theta_{\mathrm{max}}$ & $\eta_\psi$ & $\eta_\mathrm{A}$ \\
		\noalign{\smallskip} \hline \noalign{\smallskip}
		$\mathcal{A:}$ & $0$ & $0$ & $0$ & $-$ & $1$ & $1.00$ & $0.00$ & $0.00$ \\
		$\mathcal{B:}$ & $0$ & $4.98$ & $0.283$ & $\mathbb{Z}_2 \times \mathbb{Z}_2$ & $2$ & $2.6478$ & $-1.24$ & $0.319$ \\
		$\mathcal{C:}$ & $0$ & $4.06$ & $0$ & $\mathbb{Z}_2$ & $3$ & $2.00$ & $-1.00$ & $0.00$ \\
	\end{tabular}
\caption{Fixed points of the RG flow evaluated to leading order (LO) as
  described in the text.}
\label{tbl:LOFixedPoints}
\end{table}

The first line in Table \ref{tbl:LOFixedPoints} characterizes the trivial
Gaußian fixed point $\mathcal{A}$ with the mass corresponding to the only
relevant RG direction with power-counting critical exponent
$\theta_{m}=\theta_{\text{max}}=1$. The Pauli coupling is RG irrelevant at
this fixed point $\theta_{\kappa}=-1$ and the gauge coupling is marginal
$\theta_{e}=0$ with the next order given in terms of \Eqref{eq:pertbetae}
classifying the gauge coupling as marginally irrelevant; this reflects
perturbative triviality: there is no UV-complete trajectory in QED emanating
from the Gaußian fixed point that corresponds to an interacting theory at
low energies.

In the LO approximation, we find two further non-Gaußian fixed points labeled by
$\mathcal{B}$ and $\mathcal{C}$ at finite
values of the Pauli coupling $\kappa^\ast>0$, with $\mathcal{B}$ also featuring a finite
(dimensionless) mass parameter $m^\ast>0$. Taking the aforementioned discrete
symmetries into account, these fixed points occur in multiplicities according
to their nontrivial $\mathbb{Z}_2$ reflections as listed in \Eqref{eq:Z2}.

In addition, these fixed points differ by their number of relevant
directions, $n_{\text{phys}}$ counting the number of physical parameters. The largest critical exponent is listed in Table
\ref{tbl:LOFixedPoints} as $\theta_{\text{max}}$. The table also lists
the anomalous dimensions at the fixed point which both satisfy the
self-consistency criterion $|\eta_{A,\psi}|\lesssim
\mathcal{O}(1)$. It is instructive to take a closer look at the fixed
point $\mathcal{C}$ which exhibits an anomalous fermion dimension
$\eta_\psi=-1$ (and $\eta_A=0$). This value of $\eta_\psi$ corresponds
precisely to the amount required to convert the power-counting
irrelevant Pauli coupling to a marginal coupling, cf. the dimensional
scaling terms in \Eqref{eq:dtkap}; the scaling dimension of Dirac
fermions near fixed point $\mathcal{C}$ thus is similar to that of a
scalar boson near the Gaußian fixed point. In absence of further fluctuation
terms, the Pauli coupling would run logarithmically; however, the
fluctuation terms turn it into a relevant power-law running.

Finally, we observe no non-Gaußian fixed point at finite values of the gauge
coupling within the LO approximation. This is in line with the
conclusion of many literature studies that have not found a UV completion in
the theory space spanned by the standard QED bare action.

Let us now turn to the fixed-point analysis of the full truncation
without any further approximation.
In fact, we again find the same set of fixed points, see Table
\ref{tbl:fullFixedPoints}.

\begin{table}
	\centering
\begin{tabular}{ccccccccc}
	& $e$ & $\kappa$ & $m$ & multiplicity & $n_{\mathrm{phys}}$ & $\theta_{\mathrm{max}}$ & $\eta_\psi$ & $\eta_\mathrm{A}$ \\
	\noalign{\smallskip} \hline \noalign{\smallskip}
	$\mathcal{A:}$ & $0$ & $0$ & $0$ & $-$ & $1$ & $1.00$ & $0.00$ & $0.00$ \\
	$\mathcal{B:}$ & $0$ & $5.09$ & $0.328$ & $\mathbb{Z}_2 \times \mathbb{Z}_2$ & $2$ & $3.10$ & $-1.38$ & $0.53$ \\
	$\mathcal{C:}$ & $0$ & $3.82$ & $0$ & $\mathbb{Z}_2$ & $3$ & $2.25$ & $-1.00$ & $0.00$ \\
\end{tabular}
\caption{Fixed points of the given RG flow truncation.}
\label{tbl:fullFixedPoints}
\end{table}

The Gaußian fixed point $\mathcal{A}$, of course, remains unaffected by
the improved approximation. We observe the identical qualitative
features such as multiplicities and number of physical parameters
$n_{\text{phys}}$, with quantitative changes of our estimates for the
(nonuniversal) fixed-point values for $\mathcal{B}$ and $\mathcal{C}$ as
well as for the (universal) critical exponents. The quantitative
improvements arising from the full truncation in contrast to the LO
approximation are on the $\mathcal{O}(10\%)$ level. This is
self-consistent with the modification of the threshold functions upon
inclusion of anomalous dimensions as a consequence of higher-order
resummations.

\begin{figure}[t]
	\includegraphics[scale=0.35]{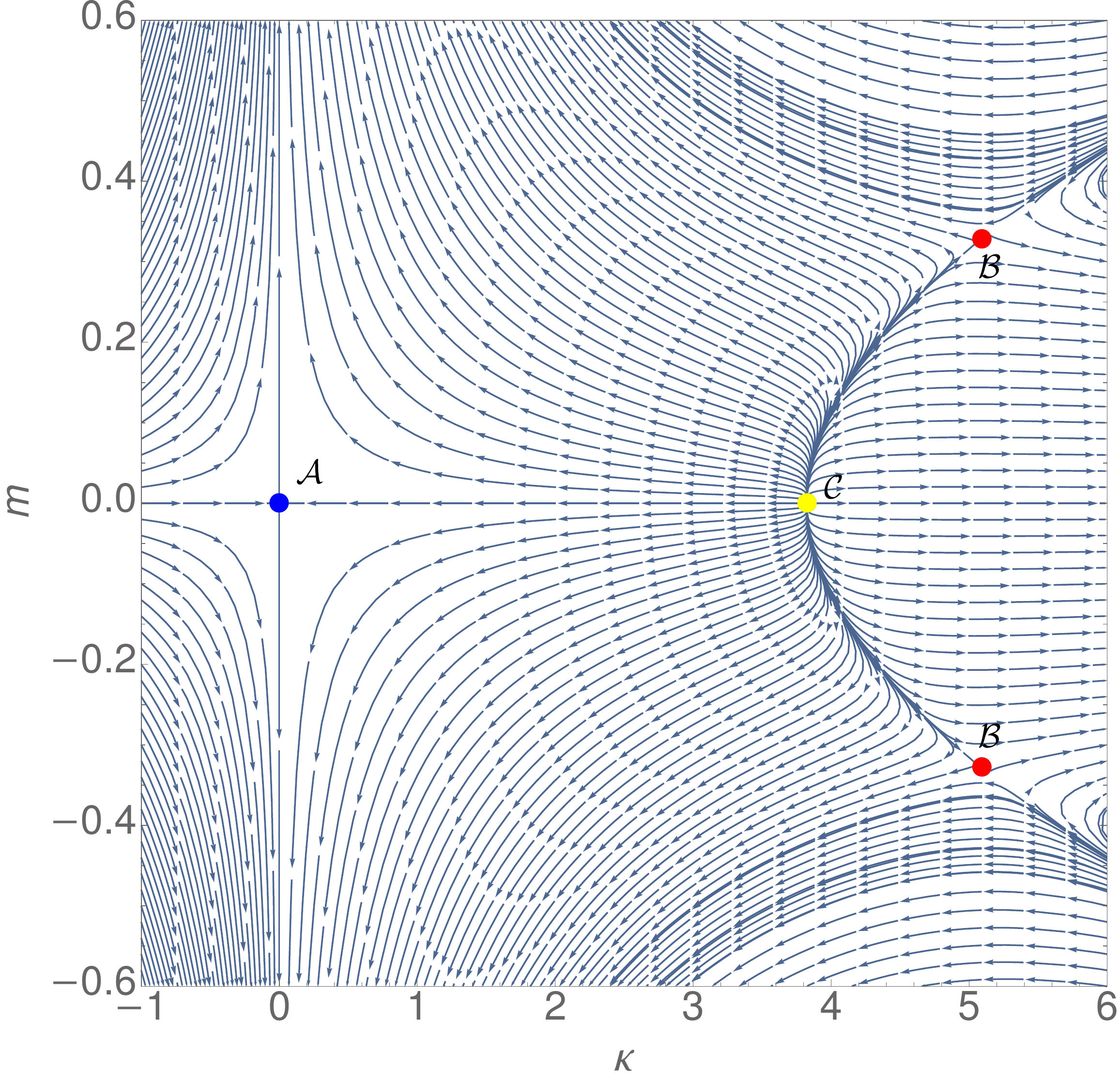}
	\caption{Phase diagram in the plane of dimensionless parameters
          $(\kappa,m)$ at $e=0$ showing the Gaußian fixed point $\mathcal{A}$,
          and the non-Gau\ss{}ian fixed points $\mathcal{B}$ (including a
          $\mathbb{Z}_2$ reflection) and $\mathcal{C}$. Arrows denote the RG
          flow towards the IR. The strongly repulsive direction at the
          Gau\ss{}ian fixed point $\mathcal{A}$ towards large values of $|m|$
          corresponds to the dimensional scaling of a mass parameter
          describing the decoupling of the massive modes. In units of a
          physical mass scale, the flow of all physical observables freezes out
          along this direction towards their long-range values. This massive
          phase can be reached from all fixed points.}
	\label{fig:StreamPlotAtEEqualTo0}
\end{figure}

The location of the fixed points and the corresponding phase diagram in the
$(\kappa,m)$ plane for $e=0$ is displayed in
Fig.~\ref{fig:StreamPlotAtEEqualTo0} with arrows pointing towards the IR. This
figure also illustrates that fixed point $\mathcal{C}$ is fully IR repulsive in
this plane, whereas $\mathcal{A}$ and $\mathcal{B}$ exhibit one attractive
direction visible in this projection. Fixed points $\mathcal{B}$ and
$\mathcal{C}$ are both IR repulsive also in the direction of the coupling $e$
(not shown in the figure), whereas $\mathcal{A}$ is marginally attractive as
dictated by \Eqref{eq:pertbetae}. Apart from the separatrices, all
trajectories emanating from the non-Gau\ss{}ian fixed points $\mathcal{B}$ and
$\mathcal{C}$ towards the region of smaller Pauli coupling eventually approach
the basin of attraction at $|m|\to \infty$ (corresponding to the formal fixed
point of the free photon gas and non-propagating electrons). This scaling
of the dimensionless mass parameter corresponds to a physical electron mass
approaching a constant value. Subsequently, the flow of all physical
observables measured in units of a physical scale freezes out, and the
observables acquire their long-range values.

Incidentally, we note that the full truncation also exhibits six other non-Gaußian fixed points and their $\mathbb{Z}_2$ reflections.
However, these fixed points have large ($\ge 6$) anomalous dimensions and therefore clearly lie outside the validity regime of our approximation.
The fact, that these fixed points do not appear in the LO approximation demonstrates that they do not pass the self-consistency test of our approximation scheme.
Therefore, we identify them as artifacts of the approximation and dismiss them in our further analysis.

In summary, we have discovered two new non-Gaußian fixed points
$\mathcal{B}$ and $\mathcal{C}$ (and their corresponding
$\mathbb{Z}_2$ reflections) of the RG flow of QED in a truncated
theory space including the Pauli coupling. They can be associated with
two new QED universality classes parametrized by
$n_{\text{phys},\mathcal{B}}=2$ and $n_{\text{phys},\mathcal{C}}=3$
physical parameters. The existence of these fixed points with a finite
number of physical parameters is a prerequisite for constructing a
UV-complete asymptotically safe version of QED.

\section{Long-range properties and physical trajectories}
\label{sec:flowIR}

Let us now construct RG trajectories related to the different fixed points. A crucial question is as to whether asymptotically safe trajectories can be constructed that are in agreement with the observed QED long-range physics.

As a warm up, we first analyze the RG flow in the vicinity of the
Gaußian fixed point (perturbative QED) without paying attention to a possibly existing UV
completion. For this, we assume $e$ to be perturbatively small over a considered range of scales, say $k\in[0,\Lambda]$. The perturbative initial condition for $\kappa$ is more subtle: as $\kappa$ is irrelevant, it is tempting to assume that the initial condition $\kappa_{k=\Lambda}$ at a the high scale does not matter too much, as long as it is in the perturbative domain. However, the long-range value of $\kappa$ is related to the celebrated result of the anomalous magnetic moment, i.e., the $g$ factor of the electron; To one-loop accuracy, we have \cite{Schwinger:1948iu},
\begin{equation}
a_e:=-  \frac{4}{e}\kappa m\bigg|_{k=0}= \frac{g-2}{2}=\frac{\alpha}{2\pi}+ \mathcal{O}(\alpha^2).\label{eq:gfactor}
\end{equation}
From the standard QED computation, it is obvious that this result is independent of the electron mass, which drops out of the corresponding projection of the electron-photon vertex. Now, the initial conditions at $k=\Lambda$ have to be chosen such that they correspond to the effective action $\Gamma_{k=\Lambda}$ which we would obtain from, say, the path integral in the presence of the IR cutoff $k=\Lambda$. In particular, the bare value for $\bk\bms|_{k=\Lambda}$ is expected to be finite, since all fluctuations with momenta above $k=\Lambda$ already have to be included. As $k$ acts like a mass parameter for all modes, we anticipate that $\bk\bms|_{k=\Lambda}$ may already be close to its physical value \eqref{eq:gfactor}, since mass scales drop out of this classic result. The details may depend on the chosen regulator.

This necessity of choosing ``loop-improved'' initial conditions
\cite{Ellwanger:1996wy} also becomes visible in the flow equations. In fact,
the flow of the dimensionless combination
\begin{equation}
  \partial_t (\kappa m)= \kappa \partial_t m + m \partial_t\kappa,\label{eq:dtkapm}
\end{equation}
shows features of a marginal coupling as the dimensional scaling terms drop out. In the perturbative domain, this flow agrees with that of  $\partial_t (\bk\bms)$ up to subleading anomalous-dimension terms. 
While the perturbative  flow is characterized by the standard log-like running of $e$ and the power-counting dimensional running of $m=\bms/k$ (with small perturbative corrections), the flow of $\kappa$ can be characterized by \Eqref{eq:dtkapm}. Anticipating $\kappa\sim e^3$ in line with \Eqref{eq:gfactor} and dropping higher-order terms, we find to leading perturbative order:
\begin{eqnarray}
  \partial_t(\kappa m) &=& v_4 e^3 m^2 [l_4^{(1,\text{B},\tilde{\text{F}}_1,\text{F})}(0,m^2,m^2)\nonumber\\
                       &&  \qquad\qquad -l_4^{(1,\text{B},{\text{F}}_1,\tilde{\text{F}})}(0,m^2,m^2)]\nonumber \\
                       && + \mathcal{O}(\kappa^3,e\kappa^2,e^2\kappa). \label{eq:pertkap}
\end{eqnarray}

Integrating this flow equation to leading order for a fixed $e$ and $\bms$ from $k=0$ to $\Lambda$, we find in the limit $\Lambda\gg\bms$ using the linear regulator:
\begin{eqnarray}
  \bk\bms|_{k=0}\simeq  \kappa m|_{k=0} &=& \kappa m|_{k=\Lambda}- \int_0^{\infty} \frac{dk}{k}\,\, \text{RHS of \Eqref{eq:pertkap}} \nonumber\\
                                        &=&\kappa m|_{k=\Lambda} -  \frac{e^3}{32 \pi^2}\, \frac{1}{6} \frac{\bms^2}{\Lambda^2}. \label{eq:gflow}
\end{eqnarray}
We observe that the flow in our current massive regularization scheme does
not induce a significant running of $\kappa m$ for $\Lambda\gg \bms$. This
confirms our expectation that the proper description of the anomalous magnetic
moment $a_e$ of the electron is essentially encoded in the boundary condition for
$\Gamma_k$ at $k=\Lambda$. Note that this boundary condition is not an
independent parameter of the theory but can be worked out from a standard
perturbative loop computation upon inclusion of the regulator term. In
practice, we fix the physical flow such that $\kappa m|_{k=0}$ corresponds to
the observed experimental value for $a_e$, see below.

Let us now turn to a discussion of the long-range properties of the system in the universality classes defined by the nontrivial fixed points $\mathcal{B}$ and $\mathcal{C}$. In contrast to the Gau\ss{}ian fixed point $\mathcal{A}$, the fixed-points  $\mathcal{B}$ and $\mathcal{C}$ allow for the construction of UV-complete RG trajectories. However, UV completeness does not guarantee that these universality classes exhibit a proper QED-type long-range behavior.
For this, it is of central interest whether one can find an RG trajectory
connecting the fixed point regimes in the UV with physical long-range behavior
defined by the IR values for all couplings. The number of relevant directions
$n_{\text{phys}}$ defines the dimensionality of the set of UV-complete RG trajectories emanating from the fixed point.

Let us start with fixed point $\mathcal{B}$ with  $n_{\text{phys},\mathcal{B}}=2$ relevant directions and critical exponents ($\theta_1=\theta_{\text{max}}=3.10$, $\theta_2=2.13$, $\theta_3=-0.81$). This implies that if we fix two parameters out of our set of couplings $(e,\kappa,m)$ the third one is a definite prediction of the universality class. In practice, we fix one parameter such that $e\approx 0.3$ in the IR corresponding to the physical value of the coupling $\alpha\simeq 1/137$. The second parameter is implicitly chosen by initiating the flow at some scale $\Lambda$ in the vicinity of the fixed point. This scale $\Lambda$ can then be expressed in terms of the resulting dimensionful electron mass $\bar{m}_e=m k|_{k\to 0}$ which defines our physical mass units. The long-range Pauli coupling $\kappa$ is then a prediction of the universality class. As an interesting subtlety, there is not just one RG trajectory, but there are actually two corresponding to a relative sign choice between our IR condition $e\simeq + 0.3$ and the discrete $\mathbb{Z}_2$ symmetries. These two trajectories are physically distinct as they go along with a different sign for the correction to the anomalous magnetic moment.\footnote{Another way of phrasing this subtlety is that if we consider flows emanating from fixed point $\mathcal{B}$ as listed in Tab.~\ref{tbl:fullFixedPoints}, we have the choice of flowing towards $e\approx 0.3$ or $e\approx -0.3$ which are both compatible with $\alpha\simeq 1/137$.}
These two trajectories correspond to distinct tangent vectors to the flow at
the corresponding $\mathbb{Z}_2$ reflections of the fixed point $\mathcal{B}$
as listed in Tab.~\ref{tbl:MassiveFPIRFlow}. Our corresponding long-range
prediction for the anomalous magnetic moment $a_e$ of the electron by following the flow from the fixed points towards the deep infrared is also listed in this table.

\begin{table}
	\centering
	\begin{tabular}{ccc}
		UV fixed point & $a_e=-4 \frac{\kappa m}{e}$ in the IR\\
		\noalign{\smallskip} \hline \noalign{\smallskip}
		$(0, 5.09, 0.328)$ & $\approx -18.55$ \\
		$(0, -5.09, 0.328)$ & $\approx 14.01$
	\end{tabular}
	\caption{First two columns: UV conditions for UV-complete RG trajectories emanating from two $\mathbb{Z}_2$ copies of fixed point $\mathcal{B}$ connecting to a long-range coupling of $e \approx 0.3$, i.e, $\alpha\simeq 1/137$. Third column: long-range prediction for the anomalous magnetic moment.}
	\label{tbl:MassiveFPIRFlow}
\end{table}

As is obvious, these predictions do clearly not match with the physical value
\begin{equation}
a_e=  -4 \frac{\kappa m}{e} \approx 0.00116.
  \label{eq:kappaBound}
\end{equation}
(For instance, the trajectory ending up with $a_e\approx -18.55$ corresponds to the separatrix emanating from fixed
  point $\mathcal{B}$ in the upper half-plane of
  Fig. \ref{fig:StreamPlotAtEEqualTo0} and then running towards $m\to\infty$
  at finite $e$.) 
We conclude that physical QED is not in the universality class of fixed point $\mathcal{B}$. Though this universality class would not be plagued by a Landau-pole problem and potentially represent a consistent quantum field theory at all scales, its long-range properties would be rather unusual: since the quantum corrections to the magnetic moment even overwhelm the Dirac value of $g=2$, strong-magnetic fields are likely to induce tachyonic modes in the spectrum of the quantum-corrected Dirac operator rendering strong and spatially extended magnetic fields unstable (similar to the Nielsen-Olesen unstable mode in nonabelian gauge theories \cite{Nielsen:1978rm}). 
Still, this version of a new QED universality class is interesting as it presents an example that QED could seem asymptotically free\footnote{This estimate of the gauge coupling appearing asymptotically free may be modified in a larger truncation. Since the Pauli coupling is non-Gau\ss{}ian, it is well possible that it feeds back into the gauge coupling through higher-order operators rendering also the fixed-point value of the gauge coupling nonzero \cite{Eichhorn:2011pc,Eichhorn:2012va}. This is a rather general mechanism which has lead to the notion of a ``shifted Gau\ss{}ian fixed point'' representing a partial near-Gau\ss{}ian fixed point in a sub-set of couplings. Despite the overall non-Gau\ss{}ianity of the system, the shifted Gau\ss{}ian sub-system behaves as if it were Gau\ss{}ian.} in its gauge coupling, since $e^\ast=0$, at the expense of an asymptotically safe Pauli term. This is another example for a close connection between \textit{paramagnetic dominance} and the UV behavior of a system \cite{Nink:2012vd}.

We finally study the universality class corresponding to fixed point $\mathcal{C}$ where only the Pauli coupling acquires a nonzero value $|\kappa|  \approx 3.82$.
This fixed point features three relevant directions, $n_{\text{phys},\mathcal{C}}=3$, and hence we expect $\mathcal{C}$ to have open neighborhoods in the $(e, \kappa, m)$ space as parts of its basin of attraction. Whether or not there exist RG trajectories emanating from $\mathcal{C}$ that are compatible with the physical QED long-range properties still remains a quantitative question to be studied. Even though $n_{\text{phys},\mathcal{C}}=3$ agrees with the number of physical parameters that we wish to match, there is a priori no reason why the physical domain belongs to the ``IR window'' of such a fixed point. 

A straightforward construction starting at a UV scale in the vicinity of the fixed point -- though possible in principle -- is numerically challenging, since the large critical exponents ($\theta_1=\theta_{\text{max}}=2.25$, $\theta_2=1.79$, $\theta_3=0.413$) indicate that a substantial amount of fine-tuning of the initial conditions would be necessary to yield specific IR values. This numerical issue can be circumvented by constructing the flow from the IR towards the UV, since the fixed point is fully UV attractive in the present truncation. 

In practice, we initiate the flow close to our physical IR boundary conditions: e.g., at the mass threshold scale defined by
$k = \Lambda_m$ where $m = 1$ and the couplings being close to their IR values $e \approx 0.3$ and $\kappa \approx \kappa_{\mathrm{phys}}$ satisfying \Eqref{eq:kappaBound}.
For $k$ towards smaller scales, the flow quickly freezes out as a consequence of the decoupling of massive electron modes.

\begin{figure}[t]
\includegraphics[scale=0.32]{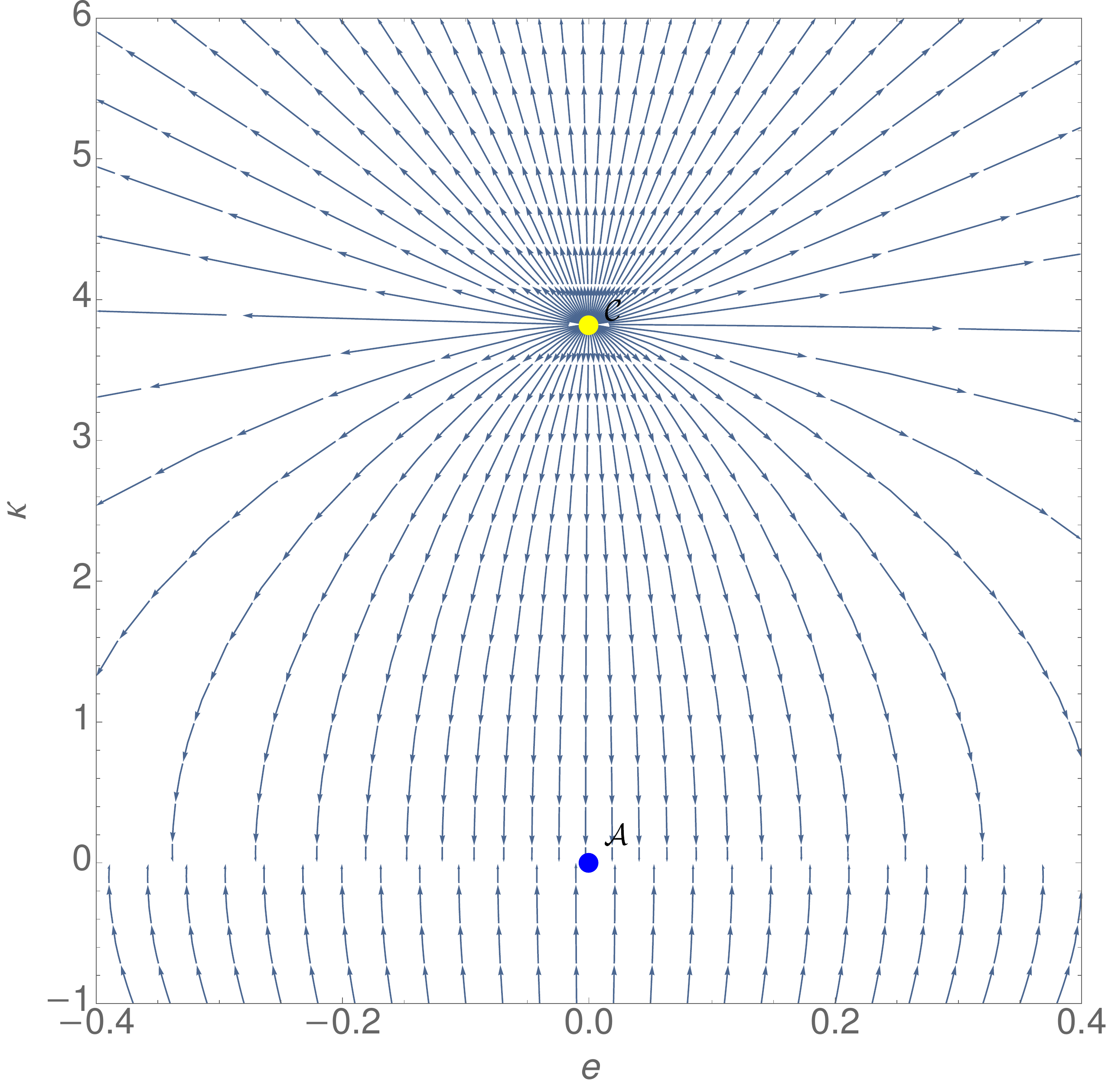}
\caption{Phase diagram in the plane of dimensionless parameters $(e,\kappa)$
  at $m=0$ showing the non-Gau\ss{}ian fixed point $\mathcal{C}$ and the
  Gau\ss{}ian fixed point $\mathcal{A}$. Flows emanating from $\mathcal{C}$
  towards smaller values of $\kappa$ span a wide range of finite gauge
  couplings $e$ in the IR, also accommodating the physical value
  $|e|\simeq0.3$. (The phase diagram near the $\kappa = 0$ axis exhibits a
  seeming singularity which is lifted by a strong flow of $m$ towards
  decoupling, implying that all trajectories freeze-out and end at $\kappa =
  0$, generically at finite values of $-4 \frac{\kappa m}{e}$.)}
\label{fig:StreamPlotAtMEqualTo0}
\end{figure}

Running the RG flow numerically from $\Lambda_m$ towards the UV, we arrive at the $\mathbb{Z}_2$ reflection of fixed point $\mathcal{C}$ with $\kappa_{\mathrm{UV}} \approx -3.82$ without any further fine-tuning.
\begin{figure}[t]
	\includegraphics[scale=0.6]{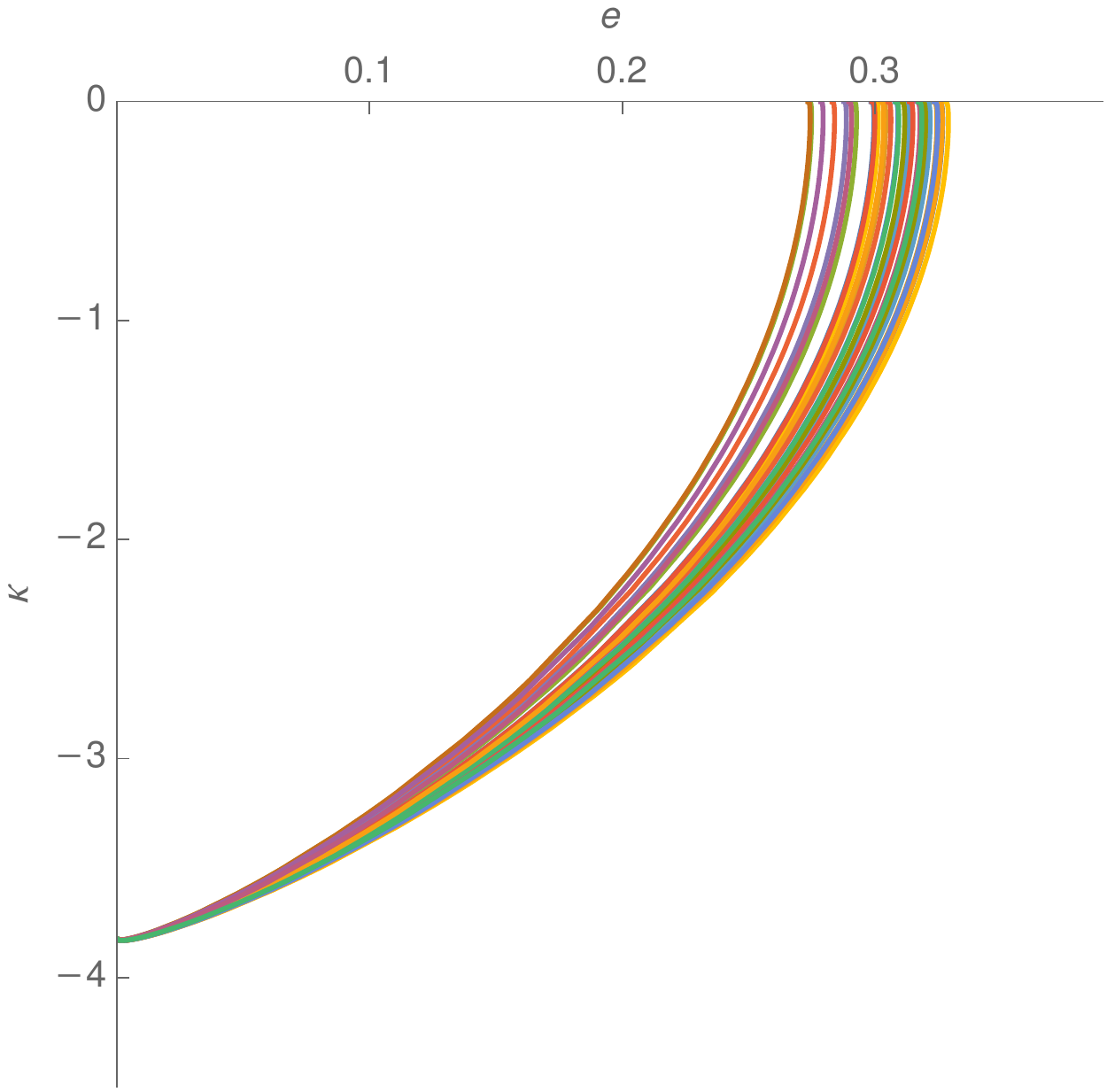}
	\caption{RG trajectories towards the UV of $30$ points close to $(0.3, \kappa_{\mathrm{phys}}, 1)$ projected to the $(e, \kappa)$ plane. All flows converge to the UV fixed point $(0, -3.82, 0)$.}
	\label{fig:UVSprayEAndKappa}
\end{figure}
We can vary the $e$ and $\kappa$ values at $\Lambda_m$ by at least $10\%$ and
still hit the same UV fixed point as is demonstrated in
Fig.~\ref{fig:UVSprayEAndKappa} where different colors correspond to
  different initial conditions in the IR; vice versa, the existence of this
  set of trajectories illustrates that, e.g., the IR value of the
  fine-structure constant $\alpha\simeq 1/137$ is not particularly
  distinguished, but merely one out of a larger interval of possible IR
  values. This is also visible in the phase diagram in the $(e,\kappa)$ plane
  at $m=0$ displayed in Fig.~\ref{fig:StreamPlotAtMEqualTo0}. A wide range of
  trajectories emanating from fixed point $\mathcal{C}$ towards smaller values
of $\kappa$ approach the small $\kappa$ region at some finite value of the
gauge coupling. At the same time, generic initial conditions lead to finite
values of the physical mass and thus to a decoupling or freeze-out behavior towards
$m \to \infty$ for the dimensionless mass parameter. This dominant IR flow
orthogonal to the $(e,\kappa)$ plane appears as a seeming singularity at
$\kappa=0$ in Fig.~\ref{fig:StreamPlotAtMEqualTo0}. In summary, the physical
IR values of
 the fine-structure constant and the anomalous magnetic moment of the electron
 can easily be accommodated in the set of trajectories emanating
   from the non-Gau\ss{}ian fixed point $\mathcal{C}$.

\begin{figure}[t]
	\includegraphics[scale=0.45]{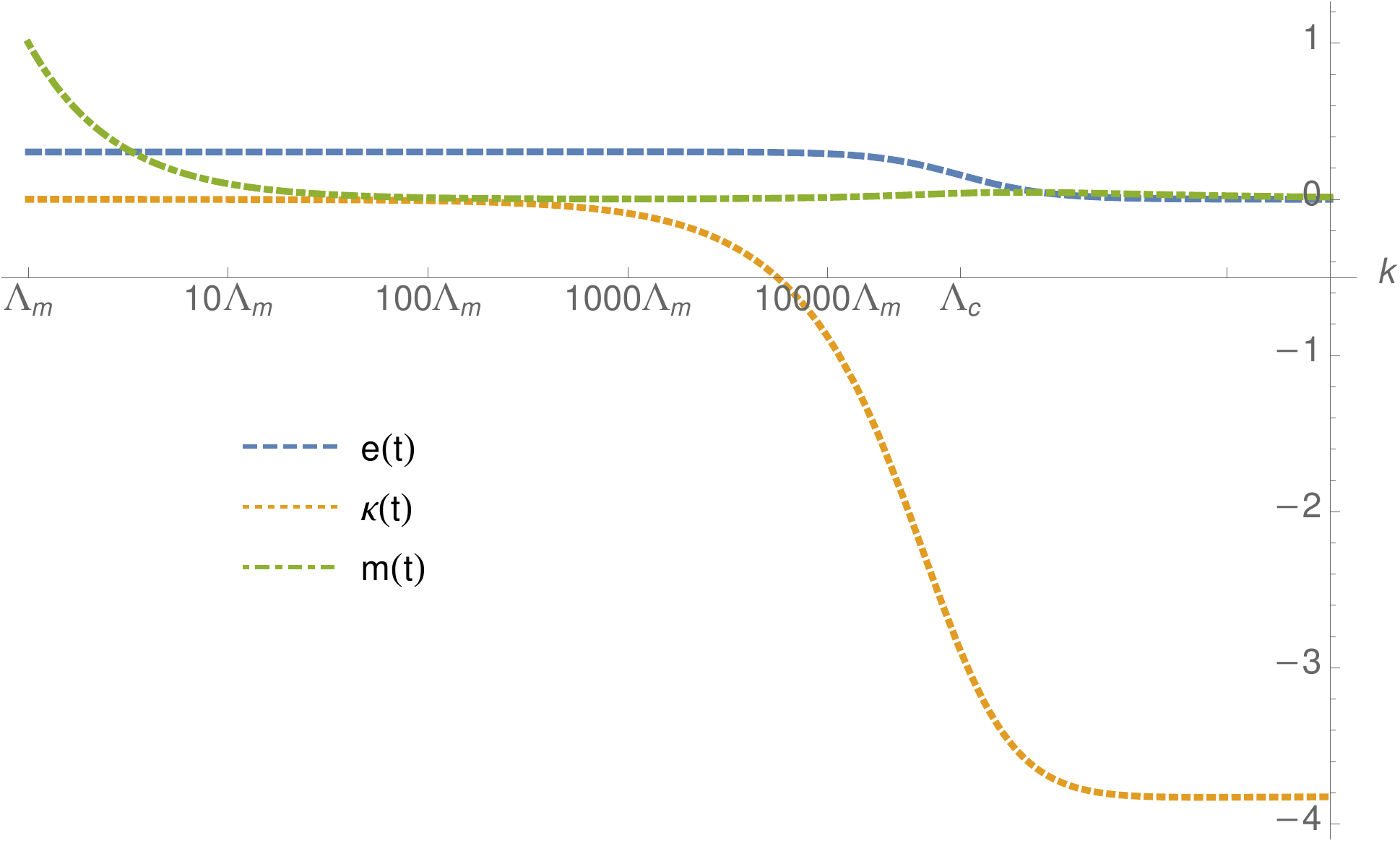}
	\caption{RG flow towards (a $\mathbb{Z}_2$ reflection of) the UV fixed point $\mathcal{C}$ $(0,-3.82,0)$. The dimensionless electron mass parameter $m$ (green) exhibits the massive decoupling in the IR near $k=\Lambda_m$, the gauge coupling (blue) is asymptotically free towards high energies, and the Pauli coupling $\kappa$ (orange) features a transition to the fixed point regime near the scale $k=\Lambda_{\text{c}}$. Note that the $k$ axis is logarithmic.}
	\label{fig:UVFlowNegativeKappa}
\end{figure}

Figure \ref{fig:UVFlowNegativeKappa} shows such a flow for intermediate values of $e$ and $\kappa$ approximately corresponding to physical IR values of the fine-structure constant and the electron anomalous magnetic moment. The dimensionless mass parameter (green line) exhibits a massive decoupling behavior in the IR near the initial scale $k=\Lambda_m$. It is interesting to observe that the flow of the gauge coupling $e$ first shows the characteristic increase towards higher energies in accordance with the perturbative running of \Eqref{eq:pertbetae} (hardly visible in the plot), but finally features asymptotic freedom with $e$ approaching its fixed point value $e^\ast\to 0$ in the deep UV. The Pauli coupling  $\kappa$ (orange) first remains perturbatively small in the IR but then undergoes a transition to its non-Gau\ss{}ian fixed-point regime.   

In Fig.~\ref{fig:UVFlowNegativeKappa}, we have introduced the scale $\Lambda_{\text{c}}$ as the scale where the flow of $e$ has its steepest slope towards asymptotic freedom. At about the same scale, $\kappa$ quickly flows towards $\kappa^\ast$.
If these flows were indeed physical (in the sense of pure QED being a fundamental theory), $\Lambda_{\text{c}}$ would denote the scale where perturbative calculations break down because of the Pauli coupling $\kappa$ becoming an RG relevant operator.
For RG flows approximately satisfying physical boundary conditions, we find
\begin{equation}
\frac{\Lambda_{\text{c}}}{\bms_e} \approx 46329
\label{eq:Lc1}
\end{equation}
in the IR limit.
In more conventional units this is equivalent to
\begin{equation}
  \Lambda_{\text{c}} \approx 46329 \cdot \bms_e
  \approx 23.67 \, \mathrm{GeV}.
  \label{eq:Lc2}
\end{equation}
By varying the IR boundary conditions for the Pauli coupling, i.e., varying the
electron anomalous magnetic moment of \Eqref{eq:kappaBound} on the
$\mathcal{O}(10\%)$ level by hand, we observe that $\Lambda_{\text{c}}$ varies
approximately linearly with $a_e$. 

It is interesting to see that this transition scale is much larger than 
the intrinsic mass scale $\bms_e$ of QED and somewhat below the 
electroweak scale.

\section{Conclusions}
\label{sec:conc}

We have studied the renormalization flow of QED in a subspace of theory space that includes the Pauli spin-field coupling. In contrast to the reduced subspace defined by perturbatively renormalizable operators, the enlarged subspace features two non-Gau\ss{}ian fixed points of the RG in addition to the Gau\ss{}ian free-field fixed point. The existence of such interacting fixed points allows for the construction of RG trajectories approaching the fixed points towards high energies thus representing UV-complete realizations of QED within the scenario of asymptotic safety. Each fixed point defines a different universality class of QED labeled by a set of critical exponents and a corresponding number of physical parameters.

One of the newly discovered fixed points (fixed point $\mathcal{C}$) allows
for the construction of UV-complete RG trajectories that can be interconnected
with the long-range physics of QED as observed in Nature. In this scenario,
the UV-failure of perturbation theory as indicated by the Landau-pole
singularity is resolved by a controlled approach of the renormalization flow
towards the fixed point with a finite value of the Pauli coupling and a
vanishing value of the gauge coupling in our approximation. In pure QED, 
we
estimate this transition to occur at a crossover scale $\Lambda_{\text{c}}$
somewhat below the electroweak scale. The RG
flow below this transition scale towards long-range physics remains
essentially perturbative. A particularity of this universality class is that
it features $n_{\text{phys}}=3$ physical parameters to be fixed. In our
considerations, we use the anomalous magnetic moment of the electron in
addition to the gauge coupling and the electron mass as additional input. In
this sense, this UV-complete version of QED has less predictive power than
perturbative QED. However, the latter has to be considered as an effective
field theory requiring the implicit assumption that all possible higher-order
operators are sufficiently small at some high scale. By contrast, QED in universality class $\mathcal{C}$ controls all further higher-order operators by virtue of the fixed point.

The other newly discovered fixed point $\mathcal{B}$ also allows for
UV-complete versions of QED fixed by only two physical parameters and thus has
the same predictive power as perturbative QED. However, our estimates of the
corresponding long-range physics feature rather large values for the anomalous
magnetic moment of the electron which are incompatible with observation.
If pure QED was a correct description of Nature, low-energy observations would
already rule out a UV completion of QED in universality class
$\mathcal{B}$.

Our estimates of the RG flow in the enlarged QED theory space are based on the functional RG which can address both perturbative as well as nonperturbative regimes. Our truncation of theory space is complete to lowest nontrivial order in a combined operator and derivative expansion. While higher-order computations will eventually be required to check the convergence of this expansion scheme, we have performed an intrinsic consistency check by quantifying the contributions of derivative operators in terms of anomalous dimensions. A comparison of leading-order to next-to-leading order results shows variations on the $\mathcal{O}(10\%)$ level for nonperturbative quantities, while qualitative results remain unchanged. At the non-Gau\ss{}ian fixed points, the anomalous dimensions become large enough to turn the perturbatively irrelevant Pauli term into a relevant operator, but remain sufficiently small preserve the ordering of operators according to their power-counting dimension apart from $\mathcal{O}(1)$ shifts. In summary, we consider our results as first evidence for an asymptotically safe realization of QED.

Our study had also been motivated by a recent analysis of the Pauli coupling and its influence on the UV-running of the QED gauge coupling within effective field theory \cite{Djukanovic:2017thn}. We confirm the conclusion of \cite{Djukanovic:2017thn} that the Landau pole can be screened by the Pauli coupling. In addition, we find that the running of the Pauli coupling itself can be UV stabilized by fluctuations leading to the existence of the fixed points. We also observe that it is important to treat the mass parameter on the same level as the couplings, since one of the fixed points occurs at a finite dimensionless mass parameter, invalidating the standard assumption of asymptotic symmetry.

The resulting scenario of asymptotically safe QED also fits into the picture
developed in \cite{Nink:2012vd}, observing that strong ultra-local paramagnetic
interactions can dominate the RG behavior of coupling flows. We hope
that our findings serve as an inspiration for searches for non-Gau\ss{}ian
fixed points in QED using other nonperturbative methods:
Within functional methods, vertex expansions offer a powerful expansion scheme; in
fact, vertex structures overlapping with the Pauli term are found to play an important
role in the strong coupling region of QCD \cite{Mitter:2014wpa}. New lattice
searches would need to go beyond the standard bare QED lattice action and
also require an explicit parametrization of the Pauli term and a
corresponding independent coupling;
for an example of asymptotic safety discovered on the lattice in a scalar
model, see \cite{Wellegehausen:2014noa}.
Studying the existence of these fixed points
would also be an interesting target for the conformal bootstrap along the
lines of \cite{Li:2020bnb}.

Whether or not the mechanisms and universality classes observed in the present work on pure QED can analogously be at work in the standard model remains to be investigated. Because of the chiral symmetry of the standard model, the analogue of the Pauli term corresponds to a dimension-6 operator also involving the Higgs field. Nevertheless, if asymptotic symmetry is not present in the UV as in the models of \cite{Gies:2013pma,Gies:2015lia,Gies:2016kkk,Gies:2018vwk,Gies:2019nij}, analogous mechanisms as revealed here in pure QED can be at work and thus pave novel ways towards an asymptotically safe completion of the standard model.




\section*{Note added}
In the present version (v3), we have corrected Eq.~\eqref{eq:dtkap} by a 
term that was missing in the previous and the published versions, yielding a 
subdominant correction to the values in Tab.~\ref{tbl:MassiveFPIRFlow} on the 
few percent level. Also the figures have been updated correspondingly though 
the corrections are hardly visible. A typo in the code has been fixed that 
changed the quantitative scale estimate in Eqs.~\eqref{eq:Lc1} and 
\eqref{eq:Lc2} substantially compared to the previous and published versions.

\acknowledgments

We are grateful to Andreas Wipf, and Luca Zambelli for valuable
discussions. Comments and suggestions by Dalibor Djukanovic, Aaron Held, Leonhard Klar, Rob Pisarski,
and Robert Shrock are thankfully acknowledged. We are particularly 
grateful to Kevin Tam for comments that lead to the improved version (v3) of 
this paper.
This work has been funded by the Deutsche Forschungsgemeinschaft (DFG) under Grant Nos. 398579334 (Gi328/9-1) and 406116891 within the Research Training Group RTG 2522/1.

\section{appendix}
\label{sec:appendix}

A widely used nomenclature for the threshold functions that parametrize the decoupling of massive modes, has already been introduced in early applications of the Wetterich equation, see, e.g., \cite{Jungnickel:1995fp}. However, the present model requires a large number of threshold functions which have not been considered so far, because of the explicit breaking of chiral symmetry and because of the momentum dependence of the Pauli coupling. We therefore suggest a more comprehensive nomenclature of threshold functions that covers all cases typically studied in the literature, as well as the new cases required by this project, and leaves room for further generalizations. We define the threshold functions used in section \ref{sec:floweq} as follows:
\begin{widetext}
\begin{eqnarray}
l_d^{([n], X_{[x_d]}^{[x_p]}, Y_{[y_d]}^{[y_p]}, ...)}(\omega_X, \omega_Y, ... ; \eta_X, \eta_Y, ...) &=&
\left(-1\right)^{1 + x_d \, x_p + y_d \, y_p + ...}
\frac{k^{- 2n - d + 2 x_p \left( 1 + x_d \right) + 2 y_p \left( 1 + y_d \right) + ...}}{4 v_d}
\label{eq:thresholdNotationDef}\\
            &&\times
               \int     \frac{\mathrm{d}^d p}{\left(2 \pi \right)^d}
               \left( p^2 \right)^n \tilde{\partial}_t
\left[ \left( \frac{\partial}{\partial p^2} \right)^{x_d} \Gprop_X(\omega_X) \right]^{x_p}
\left[ \left( \frac{\partial}{\partial p^2} \right)^{y_d} \Gprop_Y(\omega_Y) \right]^{y_p} \dots
   \nonumber
\end{eqnarray}
\end{widetext}
Here, parameters in brackets are optional and are understood to have standard defaults ($n = 0, x_d = 0, y_d=0, \dots, x_p = 1,y_p=1, \dots$). The sign conventions are such that all threshold functions are positive for finite mass parameters $\omega_{X,Y,\dots}$ and vanishing anomalous dimensions $\eta_{X,Y,\dots}$). As conventional in the literature, the modified scale derivative is understood to act on the regulator terms only, see, e.g., \cite{Berges:2000ew,Gies:2006wv,Braun:2011pp}.

Moreover, $\Gprop_X(\omega)$ denotes the inverse regularized propagator of type $X$, i.e
\begin{equation}
\begin{aligned}
\Gprop_\mathrm{B}(\omega) &= \frac{1}{P_\mathrm{B} + \omega k^2} \\
\Gprop_\mathrm{F}(\omega) &= \frac{1}{P_\mathrm{F} + \omega k^2} \\
\Gprop_{\tilde{\mathrm{F}}}(\omega) &= \frac{1 + r_\mathrm{F}}{P_\mathrm{F} + \omega k^2}
\end{aligned}
\end{equation}
where
\begin{equation}
\begin{aligned}
P_{\mathrm{B}} &= p^2 \left[ 1 + r_{\mathrm{B}}\left( \frac{p^2}{k^2} \right) \right] \\
P_{\mathrm{F}} &= p^2 \left[ 1 + r_{\mathrm{F}}\left( \frac{p^2}{k^2} \right) \right]^2 \\
\end{aligned}
\end{equation}
and $r_{\mathrm{B}}, r_{\mathrm{F}}$ are the boson and fermion regulator shape functions respectively.

Our convention covers many widely used threshold functions as well as some that have been defined for specific studies. For instance, in comparison to the notation used in \cite{Gies:2013pma}, we have the following correspondence
\begin{equation}
\begin{aligned}
l_d^{(\mathrm{B}^{n_1},\mathrm{F}^{n_2})} &= l_{n_1, n_2}^{(\mathrm{F}\mathrm{B})d} \\
l_d^{(1, \mathrm{F}_1^2)} &= m_2^{(\mathrm{F})d} \\
l_d^{(2, \tilde{\mathrm{F}}_1^2)} &= m_4^{(\mathrm{F})d} \\
l_d^{(\mathrm{B},\tilde{\mathrm{F}})} &= a_3^d.
\end{aligned}
\end{equation}
Let us finally list the explicit forms of the threshold functions as they are
needed for the present work in $d=4$, employing the linear regulator \cite{Litim:2000ci,Litim:2001up} for $r_{\mathrm{B}}$ and $r_{\mathrm{F}}$:

\begin{widetext}
\begin{equation}
\begin{aligned}
l_4^{(\mathrm{B},\tilde{\mathrm{F}})}(0,m^2) &= \frac{1}{60 \left(1 + m^2\right)^2} \left[ 
60 - 5 \eta_\psi + 5\left(4+\eta_\psi\right) m^2 -8 \eta_A \left(1+m^2 \right)
\right] \\
l_4^{(1,\mathrm{B},\tilde{\mathrm{F}}_1)}(0,m^2) &= \frac{1}{30 \left(1+m^2 \right)^2} \left[
- 2 \eta_A \left(1+m^2 \right) + 10 \left(3 - m^2 \right) -5 \eta_\psi \left( 1 - m^2 \right)
\right] \\
l_4^{(1,\mathrm{B},\tilde{\mathrm{F}})}(0,m^2) &= \frac{1}{210 \left(1+m^2 \right)^2} \left[
7 \eta_\psi \left(-1 + m^2 \right) - 12 \eta_A \left( 1 + m^2 \right) + 42 \left(3 +m^2 \right)
\right] \\
l_4^{(2,\mathrm{B},\tilde{\mathrm{F}}_1)}(0,m^2) &= \frac{1}{70 \left(1+m^2 \right)^2} \left[
- 2 \eta_A \left(1+m^2 \right) + 28 \left( 2 - m^2 \right) -7 \eta_\psi \left( 1 - m^2 \right)
\right] \\
l_4^{(1,\mathrm{B},\mathrm{F}_1})(0,m^2) &= \frac{5 - \eta_\psi}{5 \left(1 + m^2 \right)^2} \\
l_4^{(\mathrm{F}^2)}(m^2) &= \frac{5 - \eta_\psi}{5 \left(1 + m^2 \right)^3} \\
l_4^{(1,\tilde{\mathrm{F}}^2)}(m^2) &= \frac{\left(5 - \eta_\psi\right) \left(1-m^2 \right)}{10 \left(1+m^2\right)^3} \\
l_4^{(1,\tilde{\mathrm{F}},\mathrm{F})}(0,m^2,m^2) &= \frac{\left(6-\eta_\psi\right)\left(3-m^2\right)}{30 \left(1+m^2 \right)^3} \\
l_4^{(1,\mathrm{F}_1^2)}(0,m^2) &= \frac{1}{(1+m^2)^4} \\
l_4^{(2,\tilde{\mathrm{F}}_1^2)}(0,m^2) &= \frac{1-m^2}{4 \left(1+m^2 \right)^4} \left[
4- \eta_\psi + 2m^2 - \eta_\psi m^2
\right] \\
l_4^{(\mathrm{B},\mathrm{F}^2)}(0,m^2) &= \frac{1}{60 \left(1+m^2 \right)^3} \left[
-12 \eta_\psi - 5 \eta_A \left(1+m^2\right) + 30 \left(3+ m^2 \right)
\right] \\
l_4^{(\mathrm{B},\tilde{\mathrm{F}}^2)}(0,m^2) &= \frac{1}{12 \left(1+m^2 \right)^3} \left[
24 - 4 \eta_\psi \left( 1 - m^2 \right) - 3 \eta_A \left(1+m^2 \right)
\right] \\
l_4^{(1,\mathrm{B},\mathrm{F},\tilde{\mathrm{F}})}(0,m^2,m^2) &= \frac{1}{210 \left(1+m^2 \right)^3} \left[
- 7 \eta_\psi \left(3 - m^2 \right) - 12 \eta_A \left( m^2 + 1 \right) + 42 \left( m^2 + 5 \right)
\right] \\
l_4^{(2,\mathrm{B},\tilde{\mathrm{F}}^2)}(0,m^2,m^2) &= \frac{1}{168 \left(1+m^2 \right)^3} \left[
112 - 8 \eta_\psi \left(1-m^2\right) - 7 \eta_A \left(1 + m^2 \right)
\right] \\
l_4^{(2,\mathrm{B},\mathrm{F}^2)}(0,m^2,m^2) &= \frac{1}{360 \left(1+m^2 \right)^3} \left[
-20 \eta_\psi - 9 \eta_A \left(1+m^2\right) + 90 \left(3 + m^2 \right)
\right] \\
l_4^{(1,\mathrm{B},\tilde{\mathrm{F}}_1,\mathrm{F})}(0,m^2,m^2) &= \frac{1}{60 \left(1+m^2 \right)^3} \left[
- 5 \eta_\psi \left(3 - 2 m^2\right) + 20 \left( 4-m^2 \right) - 4 \eta_A \left( 1 + m^2 \right)
\right] \\
l_4^{(1,\mathrm{B},\mathrm{F}_1,\tilde{\mathrm{F}})}(0,m^2,m^2) &= \frac{4-\eta_\psi}{4 \left(1+m^2\right)^3} \\
l_4^{(\mathrm{B},\mathrm{F},\tilde{\mathrm{F}})}(0,m^2,m^2) &= \frac{1}{60 \left(1+m^2\right)^3} \left[
- 5 \eta_\psi \left(3-m^2 \right) - 8 \eta_A \left(1+m^2 \right) + 20 \left(5+m^2 \right)
\right] \\
l_4^{(\mathrm{B},\mathrm{F}^2)}(0,m^2) &= \frac{1}{60 \left(1+m^2\right)^3} \left[
-12 \eta_\psi - 5 \eta_A \left(1+m^2 \right) + 30 \left(m^2 + 3 \right)
\right] \\
l_4^{(1,\mathrm{B},\tilde{\mathrm{F}}^2)}(0,m^2) &= \frac{1}{60 \left(1+m^2\right)^3} \left[
60 - 6 \eta_\psi \left(1-m^2\right) - 5 \eta_A \left(1+m^2 \right)
\right] \\
l_4^{(2,\mathrm{B},\mathrm{F},\tilde{\mathrm{F}}_1)}(0,m^2) &= \frac{1}{210 \left(1+m^2\right)^3} \left[
210 - 84m^2 - 6\eta_A \left(1+m^2 \right) + 7\eta_\psi \left(3m^2 - 4 \right)
\right] \\
l_4^{(1,\mathrm{B},\mathrm{F}^2)}(0,m^2) &= \frac{1}{168 \left(1+m^2\right)^3} \left[
-16 \eta_\psi - 7 \eta_A \left(1+m^2 \right) + 56 \left(3 + m^2 \right)
\right] \\
l_4^{(1,\mathrm{B},\mathrm{F})}(0,m^2) &= \frac{1}{168 \left(1+m^2\right)^2} \left[
-8 \eta_\psi - 7 \eta_A \left(1+m^2 \right) + 56 \left(m^2 + 2\right)
\right] \\
l_4^{(\mathrm{B},F)}(0,m^2) &= \frac{1}{60 \left(1+m^2\right)^2} \left[
-6 \eta_\psi - 5 \eta_A \left(1+m^2 \right) + 30 \left(2 + m^2 \right)
\right]
\end{aligned}
\end{equation}
\end{widetext}

\bibliography{bibliography}
  
\end{document}